# The key role of the Calán/Tololo project in the discovery of the accelerating Universe

Mario Hamuy
Departamento de Astronomía
Universidad de Chile

The Nobel Prize in Physics 2011 has just been awarded to three astronomers: Saul Perlmutter, Brian Schmidt, and Adam Riess, for their amazing discovery of the accelerating expansion of the Universe. It is a well-deserved honor for the two teams that revolutionized cosmology at the turn of the 21$^{st}$ century.

If you had a chance to read the official document issued by the Royal Swedish Academy of Science, then you would have seen this recognition of the contribution of the Calán/Tololo (C&T) project in the aforementioned discovery: "In the mean time, light curves of several nearby type Ia SNe were measured by the Calán/Tololo Supernova Survey led by Mario Hamuy, Mark Phillips, Nicholas Suntzeff (of the Cerro Tololo Inter-American Observatory in Chile) and Jose Maza (Universidad de Chile) [24]. This data was essential to demonstrate that type Ia SNe were useful as standard candles. Progress was made using a relation between peak brightness and fading time, shown by Mark Phillips [25], to recalibrate the SNe to a standard profile. The brighter ones grew and faded slower – the fainter ones faster, and the relation allowed to deduce the peak brightness from the time scale of the light curve. The few "abnormal" occurrences were filtered out."

I was, of course, very pleased at the Academy's honorable recognition. Without diminishing the achievement of our community's laureates, I would like to take this opportunity to elaborate on the role of the C&T project (Hamuy et al, 1993a, 1993b, 1994, 1995, 1996a, 1996b, 1996c, 1996d, Maza et al. 1994, Lira 1996). As can happen when science makes a tremendous advance, a lot of supporting work is overshadowed in the excitement. For those of you who don't know a lot about the C&T project, I would just like to shine a little light on its contributions to the most recent advances in cosmology.

The C&T project started quietly in 1989, a collaboration between astronomers from the Cerro Calán and Cerro Tololo observatories (Mario Hamuy, José Maza, Mark Phillips y Nick Suntzeff) with the goal to discover southern supernovae and study their usefulness as standard candles.

When we started the C&T project in 1989 no one could measure distances with the accuracy needed for detecting the acceleration of the Universe. This situation had changed dramatically by 1993-1994, time at which our group had made eight remarkable achievements, namely:

(1) the discovery of 29 Type Ia supernovae (a world record at the time, and succeeding where other astronomers had previously failed),

(2) the recording of the most precise light curves ever obtained at that time, thanks to the recently adopted revolutionary CCD technology in astronomy,

(3) the proof that Type Ia supernovae were not perfect standard candles,

(4) the demonstration that Phillips´ (1993) relation was qualitatively correct,

(5) the dependence of Type Ia supernova luminosities with galaxy types,

(6) the establishment of a method to correct the supernova luminosities for host-galaxy reddening (a.k.a "Lira law")

(7) the most precise calibration of the supernova luminosities at the time, and last but not least

(8) the establishment of key tools to measure distances with a precision never reached before.

C&T was a seminal work done by a key group of Chilean and American astronomers in Chile. It was truly revolutionary not only because our work invented the method and showed to others how to apply it, but because that work has directly led to two key results in cosmology.

First, Freedman et al (2001) established the widely accepted value of the Hubble constant (H0=72). That result was based on three things: 1) the distance to the LMC from the H0 project, 2) the supernova Hubble diagram, and 3) the calibration of the distances to the Type Ia host galaxies. There were other distance indicators in their paper, but it was the SNe that gave the result the weight. The other techniques were less accurate. So, half the Hubble constant work came from the HST data (3) and the other half from 36 nearby supernovae [26 from C&T and 10 from the CfA follow-up program (Riess et al. 1999)] (2).

Secondly, it was the C&T project that inspired Schmidt and Suntzeff to initiate the High-Z Team in 1994. In the subsequent discovery of the acceleration of the Universe by the two rival collaborations, headed by Saul Perlmutter and Brian Schmidt (Riess et al. 1998, Perlmutter et al. 1999), the C&T light curves represent half of the measurements.

As Allan Sandage wrote in 1970, practical cosmology is about the "Search for Two Numbers", the Hubble constant ($H_0$) and the deceleration parameter ($q_0$). C&T provided half of the necessary data for $H_0$ and also half of the data for $q_0$. We now think we know what those two values are, and I am proud that C&T provided half of the necessary data for $H_0$ and also half of the data for $q_0$.

**Acknowledgments**

The C&T was hosted by the Cerro Calán Observatory of the Astronomy Department of Universidad de Chile and by the Cerro Tololo Inter-American Observatory.

The team members were: Roberto Antezana, Roberto Avilés, Ricardo Covarrubias, Luis González, Paulina Lira, José Maza, César Muena, Mark Phillips, Robert Schommer (RIP), Chris Smith, Nick Suntzeff, Geraldo Valladares, Lisa Wells y Marina Wischnjewsky (RIP). The achievements of the C&T project were only possible thanks to their tireless dedication.

The C&T project was funded through the support of Universidad de Chile, the Cerro Tololo Inter-American Observatory and the Chilean Fondecyt program through grant 1920312("Search for Supernovae").